\documentclass[aps,pre,twocolumn,showpacs]{revtex4-1}
\usepackage{graphicx}
\usepackage{amsmath}
\usepackage{amssymb}
\usepackage{color}
\usepackage{psfrag}

\begin{document}
\title{Reconstruction of chaotic neural network from observed firing rates}
\author{A. Pikovsky}
\affiliation{Institute for Physics and Astronomy, 
University of Potsdam, Karl-Liebknecht-Str. 24/25, 14476 Potsdam-Golm, Germany}
\affiliation{Department of Control Theory, Nizhni Novgorod State University,
Gagarin Av. 23, 606950, Nizhni Novgorod, Russia}

\begin{abstract}
Randomly coupled neural fields demonstrate chaotic variation of firing rates, 
if the coupling is strong enough, as has been shown by
Sompolinsky et. al [Phys. Rev. Lett., v. 61, 259 (1988)]. We present a method for reconstruction of the coupling
matrix from the observations of the chaotic firing rates. The approach is based on the particular property
of the nonlinearity in the coupling, as the latter is determined by a sigmoidal gain function. We demonstrate
that for a large enough data set, the method gives an accurate estimation of the coupling matrix and
of other parameters of the system, including the gain function.
\end{abstract}
\date{\today}
\date{\today}

\maketitle
\section{Introduction}
Understanding connectivity of networks of coupled dynamical
units is a general problem appearing not only in physics, but also 
in ecology, epidemiology, genetic regulation, and climate dynamics
(see, e.g., Refs.~\cite{Deza_etal-15,*Siguhara_atel-12,*Tomovski-Kocarev-15,*Li_etal-11}).
A particularly important application field is neuroscience, where revealing
brain connectivity is a topic of hot current 
interest~\cite{Boly_et_al-12,*Pastrana-13,*Sporns-13}. A general goal here
is to reconstruct the interactions between the nodes basing on the observations
of neurophysiological signals , e.g., on the 
multichannel EEG or MEG measurements~(see 
Refs.~\cite{Smirnov_et_al-07,*Skudlarski_etal-08,*Chicharro-Andrzejak-Ledberg-11,*Yu-Parliz-11} 
and recent review \cite{Lehnertz-11}).

Many methods developed here are based on cross-correlations
and mutual information analysis, applicable to general stochastic 
processes~\cite{Schelter-Timmer-Eichler-09,*Andrzejak-Kreuz-11,*Rubido_etal-14,*Tirabassi_etal-15}.
However, if the data belong to a special class of processes with a known structure of the dynamical laws, 
much better reconstruction of connectivity can be achieved by use of
special methods developed for such a particular class. For example, if the signals can 
be considered as those from self-sustained oscillating units,
powerful methods of analysis based on the phase dynamics equations have
been developed~\cite{Kralemann-Pikovsky-Rosenblum-11,*Kralemann-Pikovsky-Rosenblum-14}.

In this paper we suggest a method for network reconstruction under assumption
that the observed chaotic neural fields are firing rates, interacting according
to a widely accepted model for neural field dynamics (see Section~\ref{sec:nnm} below). 
Each field is influenced by many others, what makes the problem of reconstruction 
non-trivial. On the other hand, the local dynamics is governed by a scalar
differential equation, structure of which is rather simple, what makes the whole
problem tractable. Below we assume only the knowledge of a general structure
of the underlying dynamical equations, but not particular regularity: thus
our approach generalizes that of Ref.~\cite{Levnajic-Pikovsky-14}, where knowledge of
the functions determining the dynamics has been assumed. Our method
is analogous to the approach of reconstruction of a network
of time-delayed units, suggested and applied to experimental data in 
Ref.~\cite{Sysoev_etal-14}.

The paper is organized as follows. We introduce the neural network model and demonstrate
its chaotic behavior in Section~\ref{sec:nnm}. The method for reconstruction of
the connectivity and its application to the network introduced in Sec.~\ref{sec:nnm} is
described in
Section~\ref{sec:rcm}. Further possible extensions are discussed in Conclusion.

\section{Neural Network Model and its Dynamics}
\label{sec:nnm}
In this paper we focus on reconstruction of the network structure that governs neural fields
in the firing rates formulation, one of the basic models
in computational neuroscience (see Refs.~\cite{Hoppensteadt-Izhikevich-97,*Bressloff-12}, here we particularly 
follow book~\cite{Ermentrout-Terman-10}). Each of $n$ nodes is characterized by its time-depending
firing rate $x_j(t)$, which
evolves depending on inputs from other nodes according to a system of ordinary differential equations
\begin{equation}
\tau_j \frac{dx_j}{dt}+x_j=F_j\left(\sum_{k=1}^n w_{jk}x_k\right), \quad j=1,\ldots,n\;.
\label{eq:netw}
\end{equation}
Here $\tau_j$ is the time constant of relaxation of the field at node $j$, and $F_j$ are
gain functions at the nodes. The network is determined by the $n\times n$ coupling
matrix $w_{jk}$. As has been shown in Ref.~\cite{Sompolinsky88}, at large enough coupling
such a network demonstrates chaos, and this is a state which allows one for reconstruction
of
the network matrix $w_{jk}$ from the observations $x_j(t)$, as described below.

\begin{figure}[!hbt]
\centering
\psfrag{xlabel}[c][c]{time $t$}
\psfrag{ylabel}[c][c]{fields $x_i(t)$}
\includegraphics[width=0.9\columnwidth]{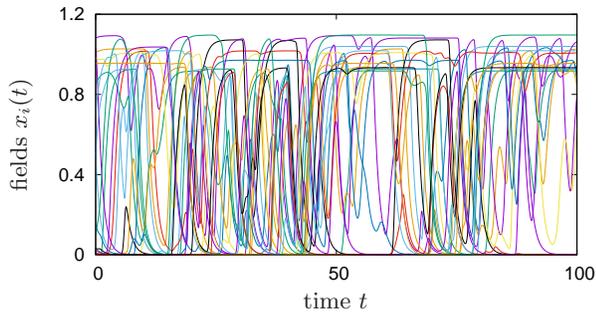}
\caption{(color online) Example of chaotic neural fields (first 20 fields $x_i(t)$, 
for $k=1,\ldots,20$ are depicted with 
different colors).}
\label{fig:fields}
\end{figure}

We illustrate a chaotic state for the following set of parameters: $n=100$; $1-\tau^0<\tau_j<1+\tau^0$
are random numbers taken from a uniform distribution with $\tau^0=0.1$. Functions $F_j$ 
have the same form but different
amplitudes: $F_j(u)=\alpha_j /[1+\exp(-u-\rho_j)]$, where $1-\alpha^0<\alpha_j<1+\alpha^0$ are 
random numbers taken from
a uniform distribution with $\alpha^0=0.1$. The links $w_{ij}$ are non-zero with probability $p_c=0.15$ (thus, 
the connections are relatively sparse),
their values are taken from a normal distribution $w_{ij}=J\cdot N(0,1)$ with $J=8$. Finally,
$\rho_i=\eta_i-0.5\sum_j w_{ij}$, where $\eta_i$ is taken from a normal distribution $N(0,1)$. Fig.~\ref{fig:fields}
shows the first 20 chaotic fields $x_j(t)$, for a realization of parameters. This chaotic state is used below for 
illustration of the reconstruction method.

\section{Reconstruction of the Connectivity Matrix}
\label{sec:rcm}
\subsection{Method of Reconstruction}
Suppose one observes time series of all variables $\vec{x}(t)$ governed by Eq.~(\ref{eq:netw}). 
The problem is to reconstruct the coupling matrix $w$ from these observations. We notice that
the functions $F_j$ and parameters $\tau_j$ are unknown and are generally different. 
We will see that
the reconstruction method allows one to reveal these quantities as well. 

The main idea is to use monotonicity of the functions $F$, which we do not 
need to know explicitly. For illustration and to simplify notations, we discuss below only 
reconstruction of the function $F_1$, of the parameter parameter $\tau_1$,
and of the coupling constants  $w_{1j}$,
all other quantities can be found similarly. We denote the row of the coupling constants
as a vector $\vec{c}$, where $c_j=w_{1j}$.

Suppose first that parameter $\tau_1$ is known.
Let us select all those points from the time series, for which $\tau_1\dot{x}_1+x_1$ lies in a small neighborhood
of a given value $y$. Let us denote the corresponding times as $t_1,t_2,\ldots,t_{m+1}$.
Let us take vectors $\vec{x}(t_k)$, $k=1,\ldots,m+1$ at these moments of time. Then, for all these vectors
\[
F_1(\vec{c}\cdot\vec{x}(t_k))\approx y\;.
\]
This means, because function $F_1$ is one-to-one, that
\begin{equation}
\vec{c}\cdot\vec{x}(t_k)\approx \vec{c}\cdot\vec{x}(t_j)\quad\text{for all}\quad k,j\;.
\label{eqc}
\end{equation}
Using the differences
\[
\vec{z}(k)=\vec{x}(t_{k+1})-\vec{x}(t_{k}),\quad k=1,\ldots,m\;,
\]
we can rewrite (\ref{eqc}) as
\begin{equation}
\vec{z}(k)\cdot \vec{c}=0\;.
\label{eq:sys1}
\end{equation}
We need to find $\vec{c}$ from this set of equations. One can see that system \eqref{eq:sys1}
does not depend on the choice of $y$, thus we can take all possible observed values of $y$ and obtain a large
set of vectors $\vec{z}$ that all satisfy \eqref{eq:sys1}. The whole set of $M$ these vectors
should be used for determining
the unknown coupling vector $\vec{c}$.

The formulated task is nothing else as solving 
homogeneous linear equations using Singular Value Decomposition (SVD), 
see, e.g., Ref.~\cite{Trefethen-Bau-97}.
The problem reduces to finding the null space of a $M\times n$ matrix $A$, composed 
of $M$ vectors $\vec{z}(k)$
as the rows. Once the zero singular value of $A$ is found, the corresponding entry in the 
obtained unitary matrix gives the vector $\vec{c}$ (up to normalization, which anyhow cannot be found by this 
method because the function $F_1$ is unknown).

Above we have assumed that the parameter $\tau_1$ is known.
In a realistic situation, parameter $\tau_1$ is unknown. Then the procedure above can be used for a set of values of $\tau_1$, chosen from a 
reasonable range. For each such value the minimal singular value of matrix $A$ can be found, and the proper 
$\tau_1$ should be chosen as yielding the minimum of these singular values.

The method described above is based on the simple observation, that close values of the function $F_1$ mean that
the arguments of this function are also close to each other. However, typically function $F_1$ is a sigmoidal 
function (in models often $\tanh(\cdot)$ is used), which have domains with derivative close to zero, where
the inversion is nearly singular. Therefore, the values of $y=\tau_1\dot{x}_1+x_1$ which are nearly constants should
be excluded from the analysis. Practically, we  use all the points for which $|\dot{y}|>\sigma$, with some
threshold $\sigma$. After all these points have been extracted from a time series, we just sorted them. In this way
the nearest neighbors after sorting are the closest points for which $y(t_1)\approx y(t_2)$, and the 
corresponding difference vector 
$\vec{z}=\vec{x}(t_1) -\vec{x}(t_2)$ is used to fill the matrix $A$.

\subsection{Numerical Results}
\begin{figure}[!hbt]
\centering
\psfrag{xlabel}[cc][cc]{time $t$}
\psfrag{ylabel}[cc][cc]{$\tau_1\dot{x}_1+x_1$}
\includegraphics[width=0.9\columnwidth]{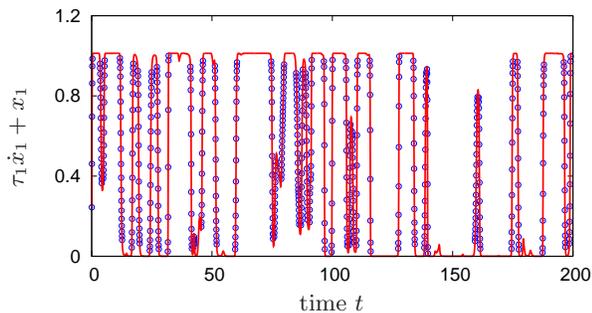}
\caption{(color online) Solid red line: $y(t)=\tau_1\dot{x}_1+x_1$  is sampled with time step $0.05$, points where
$|y|>0.3$ are shown with blue squares.}
\label{fig:rfm_der}
\end{figure}

Here we present the results of the reconstruction of coupling, for the chaotic regime presented in Fig.~\ref{fig:fields}.
Fig.~\ref{fig:rfm_der} illustrates the role of parameter $\sigma$ that discriminates tails of function $F_1$ where
its derivative is minimal. One can see that taking $\sigma=0.3$ yields points in the bulk of chaotic variations.

\begin{figure}[!hbt]
\centering
\psfrag{xlabel}[cc][cc]{$\tau$}
\psfrag{ylabel}[cc][cc]{min(SV)}
\includegraphics[width=0.9\columnwidth]{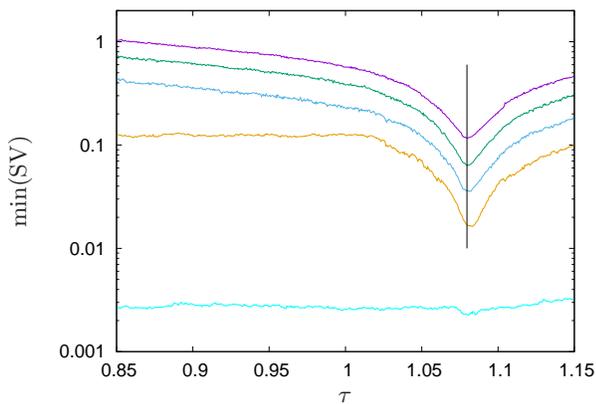}
\caption{(color online) Dependence of the minimal singular value on the parameter $\tau$ 
for different
lengths of time series (from top to bottom: total used time intervals
$2500, 1250, 500, 250, 100$). Vertical line shows the true value of $\tau$.
 }
\label{fig:rfm_tau_n}
\end{figure}

In Fig.~\ref{fig:rfm_tau_n} we show the results of calculations of the minimal singular value
for the process presented at Fig.~\ref{fig:rfm_der} with $\sigma=0.3$, in dependence on the
test values of $\tau_1$, for different total lengths of the time series. One can see that for the method
to work, the length of the time series $T$ should be large enough (in our case $T\gtrsim 250$) - 
otherwise the set of vectors $\vec{z}$ is too
small and the distances between neighbors of the sorted array of values of $y$ are too large.

\begin{figure}[!hbt]
\centering
\psfrag{xlabel}[cc][cc]{index $j$}
\psfrag{ylabel}[cc][cc]{$w_{1j},w_{1j}^r$}
\includegraphics[width=0.9\columnwidth]{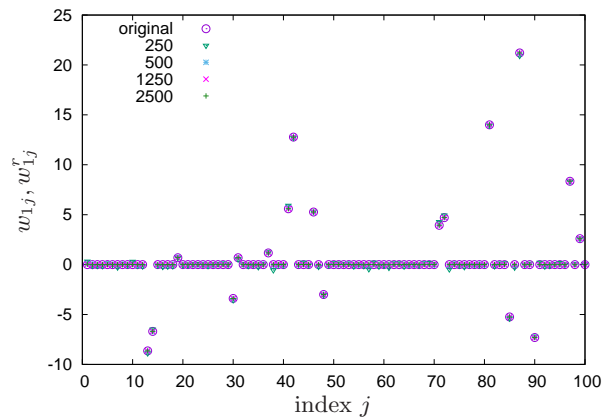}
\caption{(color online)
Original coupling constants $w_{1j}$ (circles) and the reconstructed
ones $w_{1j}^r$ for the data sets with total used time intervals
$T=2500, 1250, 500, 250$ (the corresponding markers). In these sets the 
number of data points used
for reconstruction was $8961,4174,1627,796$, respectively.}
\label{fig:rfm_res_n}
\end{figure}

Based on the analysis presented in Fig.~\ref{fig:rfm_tau_n}, in Fig.~\ref{fig:rfm_res_n} we show the results of reconstruction of the 
coupling coefficients~\footnote{Although only relative values of the coupling constants
can be reconstructed, here for clarity of comparison we normalized them by the norm of true 
coupling vector $|\vec{c}|$.}, for 4 lengths of the time series used, that demonstrate
a pronounced minimum of the singular value. The value of $\tau$
was taken from the corresponding minima. In all cases the reconstructed coupling
nearly coincides with the true one.
This proves that the accuracy
of the method is good, it allows one to infer the 
connectivity matrix from the time series. 

\begin{figure}[!hbt]
\centering
\psfrag{xlabel1}[cc][cc]{total time interval $T$}
\psfrag{ylabel1}[cc][cc]{median($|w_{1j}-w_{1j}^r|$)}
\includegraphics[width=0.7\columnwidth]{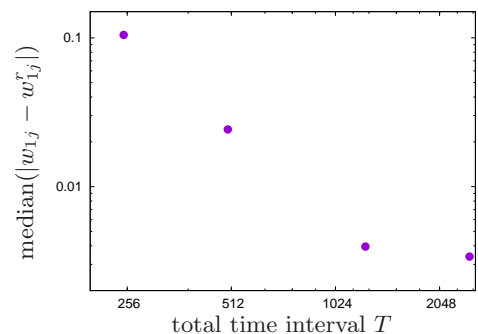}
\caption{
Median errors for the reconstruction depicted in Fig.~\ref{fig:rfm_res_n},
as functions of the total time interval used. }
\label{fig:rfm_res_n1}
\end{figure}

To characterize the accuracy (which can be hardly estimated from Fig.~\ref{fig:rfm_res_n}
as the points practically overlap), we calculated the medians of the distributions
of errors $|w_{1j}-w_{1j}^r|$, where $w_{1j}$ are coupling constants used in the
simulations (they are shown with circles in Fig.~\ref{fig:rfm_res_n}), and
$w_{1j}^r$ are reconstructed values. One can see from
Fig.~\ref{fig:rfm_res_n1} that as expected, the accuracy is 
improved if a longer time series is available.

\begin{figure}[!hbt]
\centering
\psfrag{xlabel}[cc][cc]{$u$}
\psfrag{ylabel}[cc][cc]{$F_1(u)$}
\includegraphics[width=0.4\textwidth]{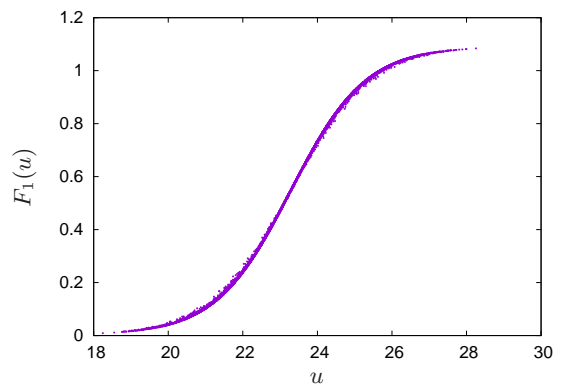}
\caption{Reconstruction of the gain function $F_1$. The same data points as in finding the coupling matrix
Fig.~\ref{fig:rfm_res_n}, with $T=2500$, are used.}
\label{fig:func}
\end{figure}

Finally, we show in Fig.~\ref{fig:func}, how the function $F_1$ is reconstructed after
the coupling constants are found.

\section{Conclusions}
In summary, we have developed a method to reconstruct the connection network behind
a collection of interacting neural fields, provided the observations of the firing
rates on the nodes are available. The method delivers the connectivity matrix, together with
the parameters characterizing node's dynamics, such as the time constant and the gain
function at each node. We have demonstrated that for a reliable reconstruction a sufficient length
of the time series is needed. In this first study we assumed a rather ideal situation where
data for all nodes are available and not contaminated by noise; exploration of the restrictions
imposed by these effects is a subject of an ongoing research.

We have formulated the method for the neural field model based on firing rates. 
There is an equivalent voltage formulation of the model where, in fact, other variables
are used~\cite{Ermentrout-Terman-10}. The approach described is not directly
suited for these variables; its corresponding generalization remains a challenging task.

\acknowledgments

We acknowledge useful discussions with V. Ponomarenko, Z. Levnajic, 
A. Daffertshofer,  and M. Rosenblum.
The work was supported by ITN COSMOS (funded by 
the European Union’s Horizon 2020 research and innovation
programme under the Marie Sklodowska-Curie grant agreement No 642563) and
by the Russian Science Foundation 
(Project No. 14-12-00811).

\def\cprime{$'$}

\end{document}